\documentclass[sigconf,nonacm]{acmart}

\setcopyright{none}

\usepackage{orcidlink}
\usepackage{booktabs}
\usepackage{multirow}

\usepackage{dblfloatfix}  % 比 stfloats 更兼容，一般更安全

\begin{document}

%%
%% The "title" command has an optional parameter,
%% allowing the author to define a "short title" to be used in page headers.
\title{PlantBGC: Transformer for Plant BGC Discovery via Label-Free Domain Adaptation and Weak Supervision}

%%
%% The "author" command and its associated commands are used to define
%% the authors and their affiliations.
%% Of note is the shared affiliation of the first two authors, and the
%% "authornote" and "authornotemark" commands
%% used to denote shared contribution to the research.
\author{Yuhan Zhao}
\orcid{0009-0006-5213-7123}
\affiliation{
  \institution{North Carolina State University}
  \department{Department of Computer Science}
  \city{Raleigh}
  \state{North Carolina}
  \country{USA}
}
\email{yzhao@ncsu.edu}

\author{Nidhi Grover}
\affiliation{
  \institution{North Carolina State University}
  \department{Department of Computer Science}
  \city{Raleigh}
  \state{North Carolina}
  \country{USA}
}
\email{nrgrover@ncsu.edu}

\author{Zhishan Guo}
\orcid{0000-0002-5967-1058}
\affiliation{
  \institution{North Carolina State University}
  \department{Department of Computer Science}
  \city{Raleigh}
  \state{North Carolina}
  \country{USA}
}
\email{zguo32@ncsu.edu}

\author{Ning Sui}
\authornote{Corresponding author}
\orcid{0000-0002-7544-7101}
\affiliation{
  \institution{North Carolina State University}
  \department{Department of Molecular and Structural Biochemistry}
  \city{Raleigh}
  \state{North Carolina}
  \country{USA}
}
\email{nsui@ncsu.edu}

%%
%% By default, the full list of authors will be used in the page
%% headers. Often, this list is too long, and will overlap
%% other information printed in the page headers. This command allows
%% the author to define a more concise list
%% of authors' names for this purpose.

%%
%% The abstract is a short summary of the work to be presented in the
%% article.
\begin{abstract}
Plant biosynthetic gene clusters (BGCs) encode specialized-metabolite pathways, yet curated plant BGC labels remain scarce, hindering supervised discovery at genome scale. Existing plant BGC mining tools are largely signature and rule-driven, and do not fully leverage recent advances in contextual representation learning for modeling long-range domain context and controlling false positives under strong domain shift. We seek an AI-assisted workflow that narrows experimental search space by transferring supervision from well-annotated microbial BGCs to plant genomes.  We present PlantBGC, representing genomes as ordered Pfam-domain sequences and learning BGC-likeness with an encoder-only Transformer trained on MIBiG microbial BGCs and adapted to plants via label-free masked language modeling. On microbial benchmarks, PlantBGC achieves token-level AUC $=0.988$ (10-fold CV) and $0.979$ (leave-class-out). On plants, adaptation improves known-BGC recovery on $n{=}34$ curated loci under strict 100\% coverage (29.4\%$\rightarrow$67.6\%), indicating more complete boundaries. GO/KEGG-derived weak supervision reduces proxy primary-like ratio by 48.40\% (GO) and 45.20\% (KEGG), with consistent per-species reductions (paired Wilcoxon $p=1.53\times10^{-5}$). Compared to \textit{plantiSMASH}, PlantBGC yields more compact loci on matched regions (median length ratio $=0.278$; 93.8\% of pairs shorter).
\end{abstract}

%%
%% The code below is generated by the tool at http://dl.acm.org/ccs.cfm.
%% Please copy and paste the code instead of the example below.
%%


%%
%% Keywords. The author(s) should pick words that accurately describe
%% the work being presented. Separate the keywords with commas.
%% A "teaser" image appears between the author and affiliation
%% information and the body of the document, and typically spans the
%% page.

\maketitle

\section{Introduction}

Natural products from bacteria, fungi, plants, and marine organisms remain foundational to medicine, agriculture, and biotechnology \citep{dias2012historical}. A substantial fraction of small molecule drugs are natural products, their derivatives, or synthetic compounds inspired by natural-product scaffolds \citep{newman2016natural}. Many bioactive molecules are synthesized by enzymes encoded by physically co-localized gene sets, known as biosynthetic gene clusters (BGCs). BGCs underlie diverse specialized metabolites, from microbial antibiotics (e.g., penicillin and erythromycin) \citep{reeves1999erythromycin} to plant therapeutics whose biosynthetic genes may also be locally clustered (e.g., monoterpene indole alkaloid loci in \textit{Catharanthus roseus} and taxane-associated arrays in \textit{Taxus} spp.) \citep{hwang2025ancient,xiong2021taxus}. This broad value motivates scalable computational strategies for BGC discovery and prioritization as genomic resources rapidly expand \citep{bauman2021genome}.

Historically, BGC discovery relied on genetic and biochemical workflows such as mutagenesis screening \citep{adrio2006genetic}, chemistry-guided isolation \citep{skinnider2015genomes}, and targeted knockout or heterologous expression for functional validation \citep{ning2022knockout}, remains definitive but costly and difficult to scale. As sequencing has become inexpensive, genome-mining systems such as antiSMASH \citep{blin2023antismash7} and ClusterFinder \citep{cimermancic2014insights} have enabled large-scale nomination using curated pHMMs and heuristic rules. However, handcrafted features and local matching assumptions can limit sensitivity to atypical or divergent clusters and struggle with long-range context, cross-species generalization, and rigorous false-positive control \citep{bauman2021genome,blin2023antismash7}.

Motivated by these limitations, machine learning and deep learning methods have been introduced for BGC detection and representation learning. Representative examples include RFBGCpred \citep{arora2025rfbgcpred}, deep sequence models such as DeepBGC and e-DeepBGC \citep{hannigan2019deepbgc,liu2022edeepbgc}, CRF-based boundary detection (GECCO) \citep{carroll2022gecco}, and self-supervised masked language modeling on domain sequences (BiGCARP) \citep{riosmartinez2023bigcarp}. Transformer-based platforms have also been proposed to unify prediction and facilitate computational design of unnatural products \citep{kawano2025transformer}. Nevertheless, the ecosystem remains predominantly microbe-centric, complicating transfer to eukaryotic systems.

Plant-derived natural products are increasingly important, yet plant genomes pose distinct challenges for BGC discovery: they are typically larger and repeat-rich, with heterogeneous gene architectures and diverse pathway organizations, leading to substantial domain shift for microbe-trained models. Plant-oriented tools such as plantiSMASH \citep{kautsar2017plantismash}, PlantClusterFinder \citep{schlapfer2017plantclusterfinder}, and PhytoClust \citep{topfer2017phytoclust} incorporate plant biological priors but remain signature-based. To our knowledge, no prior machine-learning or deep-learning framework has been developed specifically for plant BGC discovery. Existing plant-oriented methods are all rule- or statistics-based and depend on curated biosynthetic signatures, making direct comparison to learned plant-domain baselines currently impossible.
Moreover, despite extensive literature, only a few dozen plant BGCs currently have high-confidence, coordinate-resolved annotations that can be mapped to reference assemblies for supervised training and benchmarking \citep{zdouc2025mibig4}. Thus, transferring scalable AI genome mining capability from microbes to plants while minimizing reliance on plant supervision remains a key challenge.

To address this challenge, we propose PlantBGC, a Transformer-based framework that transfers BGC supervision from microbes to plants while minimizing reliance on plant labels. We represent genomes as ordered Pfam-domain token sequences and learn a contextual BGC-likeness scoring function with an encoder-only Transformer trained on curated microbial BGCs. We then perform label-free plant adaptation with masked language modeling (MLM) on unlabeled plant Pfam sequences, explicitly excluding any plant BGC labels. At inference, the adapted model produces gene/domain-level score tracks and aggregates strictly consecutive high-scoring coding sequences (CDS) into compact candidate BGC loci for downstream prioritization.

We evaluate PlantBGC along three axes. First, we assess biological validity under minimal plant supervision using locus-level recovery of curated known plant BGCs under increasingly strict coverage criteria, which probes boundary completeness rather than mere neighborhood overlap. Second, we quantify false-positive inflation using Gene Ontology (GO)\citep{ashburner2000go} and KEGG\citep{kanehisa2000kegg} as weak supervision to screen model predictions, and use the proxy primary-like ratio (ProxyFP) as a conservative surrogate for primary-metabolism-like false positives in the absence of comprehensive plant ground truth. Third, we compare boundary extent with plantiSMASH through overlap and compactness analyses on matched genomic regions, which directly impacts downstream validation cost.

\begin{figure*}[t] \centering \includegraphics[width=\textwidth]{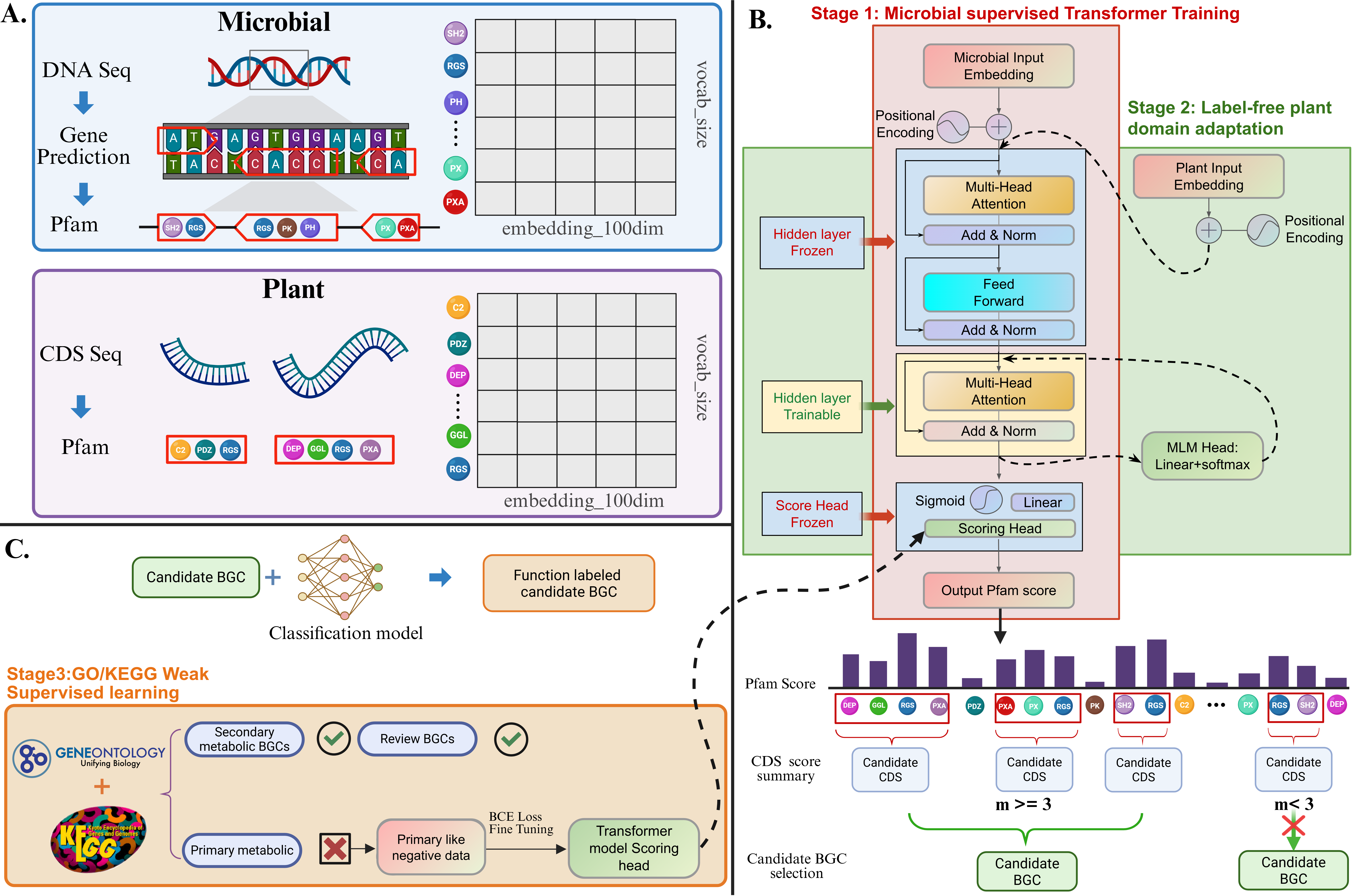} \vspace{-4mm} \caption{\textbf{Overview of PlantBGC with label transfer and weak supervision.} (\textbf{A}) Input construction for microbial and plant data. Microbial genomes and plant CDS are converted to Pfam domain tokens, each represented by a 100-dimensional Pfam2vec embedding. (\textbf{B}) Stage~1 trains a supervised microbial Transformer detector (encoder + scoring head). Stage~2 performs label-free plant domain adaptation via continued pretraining with masked language modeling (MLM) on unlabeled plant Pfam sequences while freezing the lower encoder blocks and the BGC scoring head; an auxiliary MLM head (linear + softmax) is used only during adaptation. The adapted detector outputs a score track that is pooled within each CDS and aggregated into gap-free, compact candidate BGC loci by grouping strictly consecutive high-scoring CDSs ($\tau{=}0.5$, minimum length $m{=}3$). (\textbf{C}) Stage~3 uses GO/KEGG-derived evidence for weak supervision via soft negatives; loci flagged as primary-metabolism-like are treated as soft negatives to fine-tune the Transformer scoring head with a binary cross-entropy objective.} \label{fig:overview} \end{figure*}

\vspace{-3mm}
\section{Method}\label{sec2}
\vspace{-1mm}
\subsection{Framework Overview and Problem Formulation}
\label{subsec:method_overview}

We formulate plant BGC discovery as a scoring-and-aggregation problem over CDS units represented by ordered Pfam-domain tokens: given a genome-scale CDS sequence, the goal is to output a set of compact, coordinate-resolved candidate BGC loci together with coarse functional labels. Figure~\ref{fig:overview} provides an overview of PlantBGC, which transfers supervision from curated microbial BGCs to plants via label-free adaptation and further refines plant predictions with post-hoc GO/KEGG proxy-based weak supervision.

Plant BGC discovery is fundamentally a contextual locus identification problem: specialized-metabolism loci are distinguished not only by signature domains but also by long-range and combinatorial Pfam co-occurrence patterns and boundary completeness.
Accordingly, PlantBGC is organized as a three-stage pipeline that (i) learns a transferable scoring function from abundant microbial supervision, (ii) aligns representations to plants without plant BGC labels, and (iii) calibrates against primary-metabolism-like false positives using GO/KEGG-derived weak supervision.

Specifically, we convert microbial genomes and plant CDSs into ordered Pfam-token sequences and map each token to a Pfam2vec embedding (Fig.~\ref{fig:overview}A). A Transformer detector then produces token-level BGC-likeness scores (Fig.~\ref{fig:overview}B), which are pooled to CDS-level and aggregated into gap-free loci along genomic order, yielding experimentally actionable candidate BGC loci. Finally, GO/KEGG-derived evidence provides weak supervision to reduce primary-metabolism-like false positives and to attach functional cues for prioritization (Fig.~\ref{fig:overview}C).

In Stage~1, we train the Transformer detector on microbial genomes with BGC annotations. We adopt self-attention because it can integrate long-range contextual interactions among domains and genes within genomic neighborhoods, which are often missed by local window heuristics. However, plant genomes exhibit substantial domain shift relative to microbes, while curated plant BGC labels remain scarce, making supervised training impractical. We therefore perform Stage~2 label-free domain adaptation by continued pretraining with masked language modeling (MLM) on unlabeled plant Pfam sequences to align representations across domains. At inference, token scores are pooled to CDS-level and aggregated into compact loci.
Stage~3 then uses GO/KEGG proxy evidence to provide confidence-weighted soft negatives, refining calibration without requiring plant BGC labels.

\vspace{-2mm}
\subsection{Data Processing}
\label{subsec:dataproc}
As illustrated in Fig.~\ref{fig:overview}A, we represent each genome or CDS as an ordered sequence of Pfam domain tokens, and map each token to a 100-dimensional embedding. All sequences and annotations are obtained from public resources, primarily NCBI GenBank/RefSeq and the MIBiG of experimentally characterized BGCs \citep{medema2015mibig}.

For microbes, we follow a standard genome-mining workflow: raw genomic DNA is processed by Prodigal to predict protein-coding genes \citep{hyatt2010prodigal}, and Pfam domains are assigned to predicted Open Reading Frames (ORFs) using hmmscan \citep{eddy2011hmmer,elgebali2019pfam}. 
For plants, to avoid confounding effects from intron-rich gene structures and to reduce additional ORF-calling uncertainty, we translate CDS into protein sequences and apply the same hmmscan procedure to produce plant Pfam-token sequences that are directly compatible with masked language modeling and domain-adaptation pretraining. 
We download corresponding genome annotations to map CDS units back to genomic coordinates for locus construction and evaluation.

Along each chromosome/scaffold, let $C=\{c_1,\ldots,c_N\}$ denote the ordered list of CDS units.
Each CDS $c_i$ is represented by a Pfam-token sequence $D_i=(d_{i1},\ldots,d_{iM_i})$.
All Pfam tokens are mapped to vectors trained with the skip-gram objective \citep{mikolov2013word2vec}.

We use Pfam2vec embeddings as a compact continuous representation for discrete Pfam symbols.
This design (i) exposes domain similarity and co-occurrence patterns to the model, (ii) avoids an extremely high-dimensional one-hot vocabulary, and (iii) provides a fixed input dimension required by the Transformer encoder.
We set the embedding dimension to 100 as a lightweight trade-off between expressiveness and efficiency, consistent with prior Pfam2vec practice.

\vspace{-1mm}
\subsection{Transformer Architecture and Training Stages}
\label{subsec:detector}

Figure~\ref{fig:overview}B illustrates the PlantBGC detector, an encoder-only Transformer operating on Pfam-token sequences and producing BGC-likeness scores at the token level. Transformer encoder blocks: Let $\mathbf{H}^{(0)}\in\mathbb{R}^{T\times d}$ denote the embedded token sequence of length $T$. The encoder stacks $L$ standard Transformer blocks; for block $\ell\in\{1,\ldots,L\}$, we apply multi-head self-attention (MHA) and a position-wise feed-forward network (FFN) with residual connections and layer normalization (LN): \begin{align} \tilde{\mathbf{H}}^{(\ell)} &= \mathrm{LN}\!\left(\mathbf{H}^{(\ell-1)} + \mathrm{MHA}\!\left(\mathbf{H}^{(\ell-1)}\right)\right), \\ \mathbf{H}^{(\ell)} &= \mathrm{LN}\!\left(\tilde{\mathbf{H}}^{(\ell)} + \mathrm{FFN}\!\left(\tilde{\mathbf{H}}^{(\ell)}\right)\right). \end{align} The final hidden state of token $t$ is denoted by $\mathbf{h}_t\in\mathbb{R}^{d}$. Token-level scoring head: We predict a Pfam-token-level BGC-likeness score via a linear layer followed by a sigmoid: \begin{equation} s_t=\sigma(\mathbf{w}^\top \mathbf{h}_t + b), \label{eq:token_score} \end{equation} where $s_t\in(0,1)$ is interpreted as the confidence that token $t$ belongs to a BGC-like region. Token scores are later pooled to gene-level scores and aggregated into candidate BGC loci.

\vspace{-1mm}
\subsubsection{Stage 1: Microbial Supervised Training}
\label{subsubsec:stage1}

In Stage~1 (Fig.~\ref{fig:overview}B), we train the encoder and scoring head on microbial samples. Because BGC annotations are typically available at locus (interval) resolution, we project locus boundaries to token labels: a token is labeled $y_t=1$ if the gene hosting token $t$ falls within any annotated BGC interval, and $y_t=0$ otherwise. To mitigate class imbalance, we optimize a weighted binary cross-entropy (BCE) objective: \begin{equation} \mathcal{L}_{\mathrm{sup}} = -\sum_{t}\Big(\alpha\, y_t\log s_t + (1-y_t)\log(1-s_t)\Big), \label{eq:bce_sup} \end{equation} where $\alpha$ upweights positive tokens relative to negatives. We set $\alpha$ based on the class in the training, so that positives receive proportionally stronger gradients under heavy imbalance.

\vspace{-1mm}
\subsubsection{Stage 2: label-free plant domain adaptation via MLM}
\label{subsubsec:stage2}

In Stage~2 (Fig.~\ref{fig:overview}B), plants lack sufficient high-confidence BGC labels for fully supervised training.
We therefore perform label-free domain adaptation by continued pretraining with masked language modeling (MLM) on unlabeled plant Pfam-token sequences.
We randomly select a fraction $\rho$ of token positions (typically $\rho=0.15$) as the masked set $\mathcal{M}$ and apply the standard 80/10/10 replacement rule: 80\% replaced by \texttt{[MASK]}, 10\% replaced by a random Pfam token, and 10\% kept unchanged.
An auxiliary MLM head (linear + softmax over the Pfam vocabulary) predicts the original identities at masked positions:
\begin{equation}
\mathcal{L}_{\mathrm{MLM}} = -\sum_{t\in\mathcal{M}} \log p_\theta(d_t \mid \tilde{D}),
\label{eq:mlm}
\end{equation}
where $\tilde{D}$ is the corrupted (masked) token sequence and $p_\theta(\cdot)$ is produced by the MLM head. Intuitively, MLM forces the encoder to model plant-specific Pfam co-occurrence and ordering regularities, thereby aligning the representation space from the microbial domain to the plant domain without requiring plant BGC labels.

Let $L$ denote the number of Transformer blocks in the encoder.
We freeze the lower $L_f$ blocks and update only the upper $L_u=L-L_f$ blocks during adaptation.
In addition, we freeze the token-level scoring head and update the plant embedding table $\mathbf{E}^{(p)}$ together with the MLM head.
This design preserves the BGC decision function learned under microbial supervision while allowing higher-level representations to align with plant-specific Pfam co-occurrence statistics captured by MLM.  
After adaptation, we discard the MLM head and perform inference on plants using the adapted encoder together with the frozen scoring head.
Freezing the scoring head and lower blocks explicitly anchors the microbial decision boundary during unlabeled adaptation, so Stage~2 primarily performs representation alignment rather than implicitly re-defining what constitutes a BGC. Stage~3 addresses residual calibration mismatch via proxy-guided refinement.

\vspace{-1mm}
\subsection{Candidate BGC loci aggregation and labeling}
\label{subsec:aggregation}

Figure~\ref{fig:overview}B (bottom) illustrates how PlantBGC converts token-level predictions into compact candidate BGC loci along the genomic order.
The detector outputs Pfam-token-level BGC-likeness scores.
However, experimental validation requires locus intervals rather than scattered token-level signals.
We therefore (i) pool token scores within each CDS to obtain a CDS-level score and (ii) aggregate strictly consecutive high-scoring CDSs into gap-free loci, which suppresses isolated spikes and reduces fragmented calls.

\vspace{-1mm}
\subsubsection{From token scores to candidate BGC loci}
\label{subsubsec:locus_aggregation}

Let $C=\{c_1,\ldots,c_N\}$ denote the ordered list of CDS units on a chromosome/scaffold.
Each CDS $c_i$ contains $M_i$ Pfam tokens with token scores $\{s_{ij}\}_{j=1}^{M_i}$. We summarize token scores within each unit to obtain a unit-level score: \begin{equation} S_i=\sum_{j=1}^{M_i} w_{ij}\, s_{ij}, \qquad \sum_{j=1}^{M_i} w_{ij}=1, \label{eq:unit_pooling} \end{equation} where $w_{ij}$ are normalized pooling weights (uniform by default, i.e., $w_{ij}=1/M_i$). We select high-scoring units by $S_i>\tau$ with a default threshold $\tau=0.5$. Candidate loci are then defined as maximal strictly consecutive runs of selected units (gap tolerance set to 0), producing compact, gap-free intervals consistent with the notion of physically co-localized biosynthetic genes. To suppress spurious short runs, we require a minimum run length of $m=3$ consecutive units. We define each locus interval from the start of its first unit to the end of its last unit in the genome annotation. The resulting candidate BGC loci are reported as a coordinate-ordered list along each chromosome/scaffold, with each locus represented by its genomic interval and constituent CDS.

\vspace{-1mm}
\subsubsection{Multi-label biosynthetic product classifier}
\label{subsubsec:product_classifier}

Beyond detection, we further provide coarse-grained functional labels for candidate BGC loci by predicting biosynthetic product classes and bioactivities. The goal is to offer interpretable cues for prioritization, enable summary analyses of label distributions (e.g., class skew), and complement the GO/KEGG-based evidence used in our weak-supervision training. We use random-forest models \citep{breiman2001random} as lightweight and robust classifiers suitable for sparse Pfam-derived features and limited labeled examples in MIBiG(v1.3), avoiding the need to introduce an additional deep model. We train two separate random-forest models using MIBiG annotations: one for biosynthetic product classes and one for product activities. Each BGC is represented by a binary multi-label vector indicating the presence of zero or more classes and activities. After training, we apply these classifiers to plant candidate BGC to obtain predicted profiles.

\vspace{-1mm}
\subsection{GO/KEGG weak supervision}
\label{subsec:weak_supervision}

Figure~\ref{fig:overview}C summarizes a proxy-based weak-supervision loop that calibrates plant predictions by suppressing primary-metabolism-like loci (soft negatives) while preserving high-confidence secondary/mixed candidates for prioritization. Because curated plant BGC labels are scarce, we convert GO/KEGG functional evidence into reproducible proxy labels and confidence-weighted soft targets, which provide negative feedback in Stage~3 through an additional loss term.

\vspace{-1mm}
\subsubsection{GO/KEGG functional annotation pipelines}
\label{subsubsec:gokegg_annotation}

We map Pfam domains to GO terms using the official \texttt{pfam2go} file and take the union over domains within each CDS. Let $\mathcal{T}^{\mathrm{GO}}_{\mathrm{sec}}$ and $\mathcal{T}^{\mathrm{GO}}_{\mathrm{prim}}$ denote GO term sets for secondary and primary metabolism. For each candidate locus $\ell$ with genes $\mathcal{G}_\ell$, we compute:
\begin{equation}
\begin{aligned}
\mathrm{sec\_score}^{\mathrm{GO}}_\ell
&= \sum_{g\in\mathcal{G}_\ell}\mathbb{I}\!\left[\mathrm{GO}(g)\cap\mathcal{T}^{\mathrm{GO}}_{\mathrm{sec}}\neq\emptyset\right],\\
\mathrm{pri\_score}^{\mathrm{GO}}_\ell
&= \sum_{g\in\mathcal{G}_\ell}\mathbb{I}\!\left[\mathrm{GO}(g)\cap\mathcal{T}^{\mathrm{GO}}_{\mathrm{prim}}\neq\emptyset\right].
\end{aligned}
\label{eq:go_int_scores}
\end{equation}

We then assign GO proxy labels by fixed rules:
\begin{align}
\texttt{secondary\_likely} &: \ \mathrm{sec\_score}^{\mathrm{GO}}_\ell \ge 2 \ \wedge\ \mathrm{pri\_score}^{\mathrm{GO}}_\ell = 0, \nonumber\\
\texttt{secondary\_tilt} &: \ \mathrm{sec\_score}^{\mathrm{GO}}_\ell = 1 \ \wedge\ \mathrm{pri\_score}^{\mathrm{GO}}_\ell = 0, \nonumber\\
\texttt{primary\_likely} &: \ \mathrm{pri\_score}^{\mathrm{GO}}_\ell \ge 2 \ \wedge\ \mathrm{sec\_score}^{\mathrm{GO}}_\ell = 0, \nonumber\\
\texttt{primary\_tilt} &: \ \mathrm{pri\_score}^{\mathrm{GO}}_\ell = 1 \ \wedge\ \mathrm{sec\_score}^{\mathrm{GO}}_\ell = 0, \nonumber\\
\texttt{mixed\_possible} &: \ \mathrm{sec\_score}^{\mathrm{GO}}_\ell > 0 \ \wedge\ \mathrm{pri\_score}^{\mathrm{GO}}_\ell > 0, \nonumber\\
\texttt{review} &: \ \text{otherwise}.
\label{eq:go_label_rules}
\end{align}
When $\ell$ is labeled \texttt{primary\_likely}, we additionally flag it as \texttt{unlikely\_bgc=True} for subsequent weakly supervised learning.

We assign KO identifiers to genes using BlastKOALA\footnote{BlastKOALA: \texttt{https://www.kegg.jp/blastkoala/}} or an equivalent workflow (e.g., KofamScan), and treat genes without KO as \texttt{review}. Let $\mathcal{K}^{\mathrm{KO}}_{\mathrm{sec}}$ and $\mathcal{K}^{\mathrm{KO}}_{\mathrm{prim}}$ be KO sets for specialized and primary metabolism. For each locus $\ell$:
\begin{equation}
\begin{aligned}
\mathrm{kegg\_sec}_\ell
&= \sum_{g\in\mathcal{G}_\ell}\mathbb{I}\!\left[\mathrm{KO}(g)\in \mathcal{K}^{\mathrm{KO}}_{\mathrm{sec}}\right],\\
\mathrm{kegg\_pri}_\ell
&= \sum_{g\in\mathcal{G}_\ell}\mathbb{I}\!\left[\mathrm{KO}(g)\in \mathcal{K}^{\mathrm{KO}}_{\mathrm{prim}}\right].
\end{aligned}
\label{eq:kegg_counts}
\end{equation}

We assign KEGG proxy labels: \texttt{secondary} if $\mathrm{kegg\_sec}_\ell>0$ and $\mathrm{kegg\_pri}_\ell=0$; \texttt{primary} if $\mathrm{kegg\_pri}_\ell>0$ and $\mathrm{kegg\_sec}_\ell=0$; \texttt{mixed\_possible} if both are $>0$; otherwise \texttt{review}.
In practice, \texttt{mixed\_possible} is rare and is reserved for boundary KOs.

\subsubsection{Confidence and soft-negative dataset construction}
GO and KEGG provide complementary signals; primary-like evidence may overlap but need not agree. We map proxy labels to confidence scores using the same rule for both resources: \texttt{primary\_likely}$\rightarrow 0.8$ and \texttt{primary\_tilt}$\rightarrow 0.5$, and fuse them with an agreement-boosted rule:
\begin{equation}
q_\ell = \max\!\left(q^{\mathrm{GO}}_\ell, q^{\mathrm{KEGG}}_\ell\right) + \gamma \cdot \min\!\left(q^{\mathrm{GO}}_\ell, q^{\mathrm{KEGG}}_\ell\right),
\label{eq:q_fusion}
\end{equation}
where $\gamma\in[0,1]$ (default $\gamma=0.5$). We assign token-level soft targets within locus $\ell$ by
\begin{equation}
\tilde{y}_t = 1-q_\ell.
\label{eq:soft_target_final}
\end{equation}

\subsubsection{Loss update}
Starting from the Stage~2 checkpoint, we optimize a confidence-weighted soft-label BCE objective:

\begin{equation}
\mathcal{L}_{\mathrm{soft}}
= - \sum_{\ell}\, w_\ell \sum_{t\in \mathcal{T}_\ell}
\Big(\tilde{y}_t \log s_t + (1-\tilde{y}_t)\log(1-s_t)\Big),
\label{eq:soft_bce}
\end{equation}
and combine it with microbial supervision to reduce forgetting:
\begin{equation}
\mathcal{L}_{\mathrm{stage3}} = \mathcal{L}_{\mathrm{sup}} + \lambda\, \mathcal{L}_{\mathrm{soft}}.
\label{eq:stage3_total}
\end{equation}
In Stage~3, we freeze the lower encoder blocks and update the scoring head together with the upper encoder blocks, so primary-like loci receive stronger downward gradients while uncertain cases are corrected conservatively.

\section{Experiment and Result}
\label{subsec:exp_settings}

Plant BGC discovery is a contextual locus identification problem: BGCs are characterized by co-localized gene sets whose Pfam-domain composition and co-occurrence patterns separate specialized metabolism from dense primary-metabolism regions.
Rule-based tools (e.g., plantiSMASH\citep{kautsar2017plantismash}) rely on predefined signatures and heuristics, which are effective for canonical patterns but limited in capturing higher-order context, transferring across domains, and controlling false positives under scarce plant ground truth.
We therefore organize experiments around three stage-aligned questions:
\textbf{Q1:} can a Transformer learn a stronger, more transferable BGC scoring function than prior ML baselines on curated microbial supervision (Stage~1)?
\textbf{Q2:} without plant labels, can label-free adaptation mitigate domain shift and improve locus-level recovery and boundary completeness in plants (Stage~2)?
\textbf{Q3:} after adaptation, can GO/KEGG-derived weak supervision reduce primary-metabolism-like false positives while preserving the ranking utility needed for practical prioritization (Stage~3)?
Finally, we benchmark locus compactness against plantiSMASH on matched regions to estimate validation workload. These three stages also provide a natural ablation of the proposed framework, where each stage incrementally introduces additional supervision or adaptation to address a specific limitation.

\subsection{Microbial and Plant Dataset}
\label{subsec:posneg}

We build our datasets from two public resources: the MIBiG\citep{zdouc2025mibig4} and NCBI GenBank/RefSeq.
MIBiG currently reports 2437 active curated BGC entries (in addition to pending and retired records).
Only a small fraction of these entries can be directly mapped to complete, coordinate-resolved genomic records in NCBI, we further curate microbial and plant subsets.

Microbial Positive Dataset: We map curated MIBiG entries to linked NCBI GenBank sequences and retain validated genomic loci, yielding 1420 microbial BGCs for Stage~1 supervised training.

Microbial Negative Dataset: We generate negatives using GeneSwap \citep{hannigan2019deepbgc}, which preserves local gene-length and neighborhood structure while disrupting BGC-like signals. Replacement genes are sampled from genomes screened to avoid high nucleotide similarity to known MIBiG loci (BLAST-based filtering \citep{altschul1990blast}).
For each positive segment, we create three independently swapped variants, producing a negative set three times larger than the positives.

Plant Dataset: From MIBiG dataset, we curate 17 plant species as the target domain and collect 34 known plant BGC loci for locus-level evaluation.
We use CDS sequences from NCBI RefSeq as the plant gene dataset. All predicted candidate BGC loci are ultimately represented and evaluated in the reference genome coordinate system using genome annotations.

All predicted loci and curated plant BGCs are represented as base-pair intervals on the same reference assembly for each genome using gene annotations, so that coverage and IoU are measured in a consistent coordinate system.
External tool outputs (e.g., plantiSMASH) are normalized to the same assembly version before overlap computation, ensuring that cross-method comparisons reflect boundary behavior rather than annotation-version artifacts.

\begin{table*}[t]
\centering
\small
\setlength{\tabcolsep}{5pt}
\caption{\textbf{Stage~1 microbial supervised performance.} Results are reported under 10-fold cross-validation and leave-class-out. DeepBGC and Random Forest baseline is re-trained and evaluated on our curated datasets.}
\label{tab:stage1_results}
\begin{tabular}{@{}lccccc ccccc@{}}
\toprule
\multirow{2}{*}{\textbf{Model}} &
\multicolumn{5}{c}{\textbf{10-fold CV}} &
\multicolumn{5}{c}{\textbf{Leave-class-out}} \\
\cmidrule(lr){2-6}\cmidrule(lr){7-11}
& \textbf{ROC-AUC} & \textbf{ACC} & \textbf{Prec.} & \textbf{Rec.} & \textbf{F1}
& \textbf{ROC-AUC} & \textbf{ACC} & \textbf{Prec.} & \textbf{Rec.} & \textbf{F1} \\
\midrule
PlantBGC-Transformer & \textbf{0.988} & \textbf{0.981} & \textbf{0.933} & \textbf{0.901} & \textbf{0.915} & \textbf{0.979} & \textbf{0.973} & \textbf{0.906} & \textbf{0.852} & \textbf{0.878} \\
DeepBGC-BiLSTM \citep{hannigan2019deepbgc} & 0.980 & 0.981 & 0.933 & 0.900 & 0.916 & 0.945 & 0.910 & 0.838 & 0.721 & 0.775 \\
Random Forest  & 0.923 & 0.924 & 0.786 & 0.583 & 0.669 & 0.852 & 0.862 & 0.720 & 0.565 & 0.632 \\
\bottomrule
\end{tabular}
\end{table*}

\begin{figure*}[!b]
  \centering
  \includegraphics[width=\textwidth]{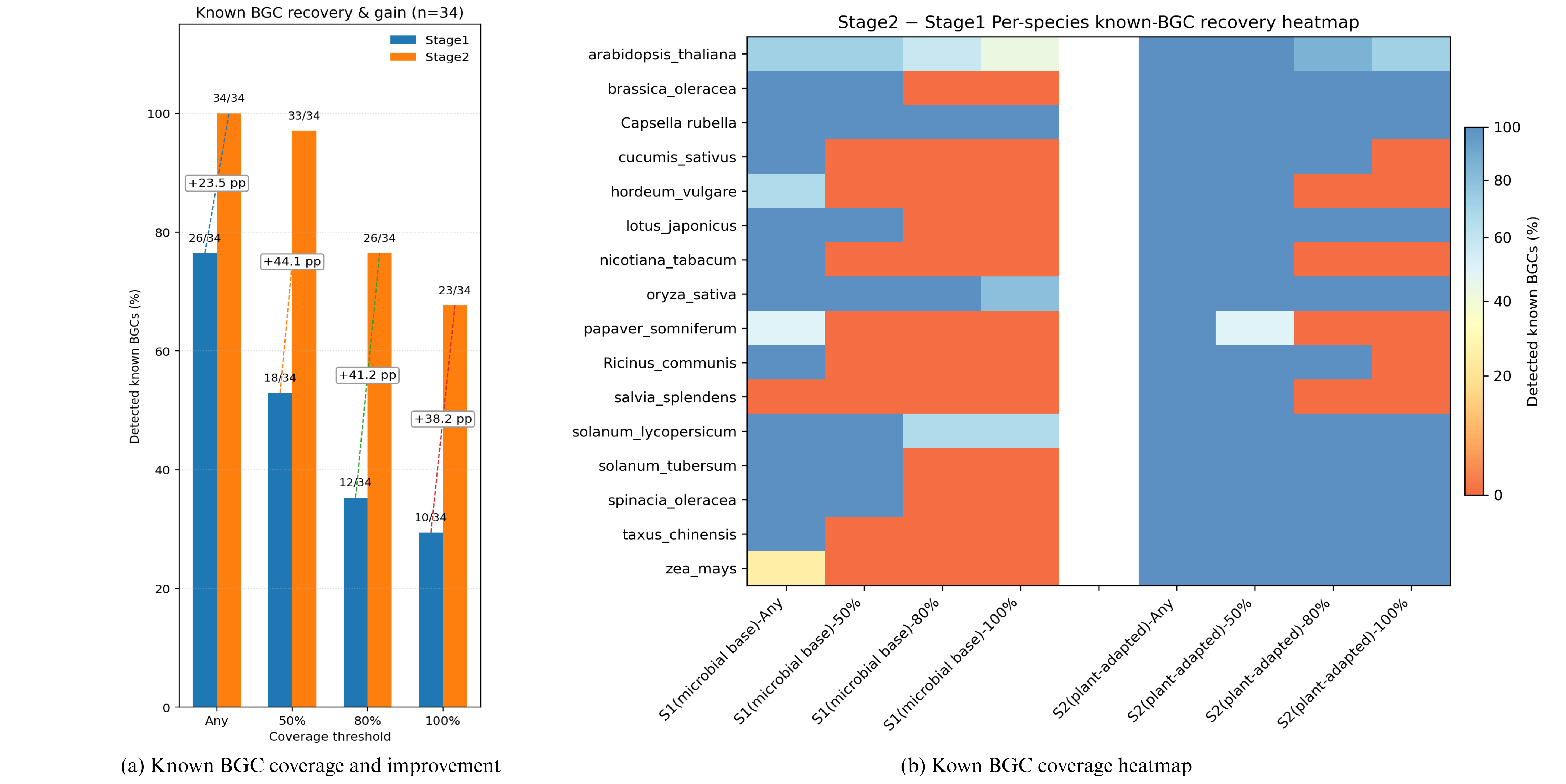}
  \vspace{-4mm}
 \caption{\textbf{Stage~2 unlabeled plant adaptation improves known-BGC recovery.}
(a) Recovery of $n{=}34$ curated plant BGCs under four coverage thresholds (Any, 50\%, 80\%, 100\%), comparing the Stage~1 microbial model and the Stage~2 plant-adapted model. Annotations indicate absolute improvement (percentage points).
(b) Per-species recovery heatmap across thresholds. Stage~1 often declines at stricter thresholds, whereas Stage~2 remains high across most species.}
  \label{fig:stage2_cov}
\end{figure*}

\vspace{-3mm}
\subsection{Stage 1: Microbial supervised Transformer learning}\label{subsec:stage1_exp}
Stage~1 addresses \textbf{Q1}: whether an encoder-only Transformer can learn a more transferable BGC scoring function from microbial supervision than established baselines. 
This stage training do not include plants, but it validates that self-attention over Pfam-domain sequences yields stronger ranking under heavy class imbalance and better generalization to unseen biosynthetic classes, which is the prerequisite for any credible cross-domain transfer in Stage~2.

We use an encoder-only Transformer and a token-level linear layer with sigmoid activation outputs BGC-likeness scores. We train with maximum sequence length 512 tokens, batch size 16,  Adam optimizer (learning rate $10^{-3}$) and weighted binary cross-entropy (BCE). We apply early stopping by monitoring validation ROC-AUC with patience 20 and minimum improvement 0.01.

We compare against following two baseline models:
\vspace{-2mm}
\begin{itemize}
\item DeepBGC\citep{hannigan2019deepbgc}: We follow the DeepBGC set and train a stateful BiLSTM token classifier on Pfam-domain sequences. The BiLSTM uses 128 hidden units and a time-distributed sigmoid head, optimized with weighted BCE and Adam (lr $=10^{-4}$) on fixed-length sequence chunks.
\item Random Forest: We train a Random Forest classifier on hand-crafted Pfam features computed per training segment and window, using class balancing and tuning hyperparameters on training folds.

\end{itemize}

We report ROC-AUC as a threshold-free measure of ranking quality under class imbalance, together with thresholded token-level metrics (ACC/Precision/Recall/F1) to quantify the operating-point trade-off between false positives and sensitivity.

Generalization is estimated with 10-fold cross-validation on paired MIBiG-derived positive segments and GeneSwap-generated negatives: in each fold we train on 9 folds and evaluate on the held-out fold, then aggregate predictions across all held-out folds.

To assess robustness to unseen biosynthetic pathway types, we perform leave-class-out (LCO) evaluation over major non-hybrid classes (e.g., PKS, NRPS, RiPP, terpene). For each held-out class $c$, we remove all positive segments of class $c$ from training, train on positives from the remaining classes, and evaluate on positives from class $c$. Since GeneSwap negatives are not class-labeled, we draw negatives from the same pool and split them into train and test using the same proportions across methods. We aggregate results across held-out classes.

Table~\ref{tab:stage1_results} shows that the Transformer achieves the strongest overall ranking quality and operating-point balance under both protocols. 
Under 10-fold CV, PlantBGC improves ROC-AUC to 0.988 and yields the highest F1 (0.915), slightly outperforming DeepBGC-BiLSTM and the Random Forest baseline. 
This indicates that the model achieves at least comparable, and in most cases stronger, in-domain learning capacity under standard evaluation. Under the complementary leave-class-out evaluation (ROC-AUC=0.979, F1=0.878), where test positives belong to biosynthetic classes excluded during training, the performance gap becomes substantially larger. 
This result suggests that self-attention over Pfam-domain sequences captures higher-order contextual cues that generalize beyond within-class interpolation, providing stronger robustness to unseen biosynthetic patterns. 
Such cross-class generalization is a key prerequisite for subsequent plant-domain transfer (Stage~2) and weakly supervised refinement (Stage~3).
Across the full ROC curves, PlantBGC consistently maintains a higher true-positive rate at the same false-positive rate than both baselines, with the largest separation observed in the leave-class-out setting.

\subsection{Stage 2: label-free plant domain adaptation via MLM}\label{subsec:stage2_exp}

Stage~2 addresses \textbf{Q2}: given a detector trained with microbial BGC supervision, whether we can improve plant performance without using any plant BGC labels. This question matters because plants exhibit substantial domain shift (gene organization and Pfam co-occurrence statistics), and only a few dozen coordinate-resolved plant BGCs exist for supervised retraining. We therefore compare (i) the Stage~1 microbial base model applied to plants, and (ii) a plant-adapted model obtained by continued pretraining on unlabeled plant Pfam sequences via masked language modeling. Because no prior ML-based plant BGC detector exists, this Stage~1 to Stage~2 comparison also serves as the fairest available plant-domain baseline: both models share the same architecture and post-processing, differing only in whether unlabeled plant adaptation is applied.

Crucially, we keep the locus aggregation identical across Stage~1 and Stage~2 so that any gain reflects representation alignment rather than post-processing. We average-pool token scores to obtain per-CDS scores, select units with $S_i>\tau$ ($\tau{=}0.5$), and merge strictly consecutive selected units (gap tolerance $=0$) into candidate loci with minimum length $m{=}3$ CDSs. All Stage~2 gains therefore reflect model differences rather than post-processing changes.

For each curated known plant BGC, we identify the best-matching predicted candidate locus by maximizing base-pair overlap.
We define coverage as the fraction of the known BGC interval covered by the best-matching prediction:
\vspace{-1mm}
\begin{equation}
\mathrm{cov}(\mathrm{pred},\mathrm{truth})=\frac{\lvert \mathrm{pred}\cap \mathrm{truth}\rvert}{\lvert \mathrm{truth}\rvert}.
\end{equation}

Recovery is evaluated at four localization stringencies: any-overlap ($\mathrm{cov}>0$), and coverage thresholds of 50\%, 80\%, and 100\% ($\mathrm{cov}\ge 0.5,0.8,1.0$).
For each threshold, we report the fraction of recovered loci among $n{=}34$ known BGCs, with 95\% bootstrap confidence intervals.

Figure~\ref{fig:stage2_cov} shows that Stage~2 unlabeled adaptation improves locus-level recovery of curated plant BGCs over the Stage~1 microbial base model.
In the pooled analysis (Figure~\ref{fig:stage2_cov}a--b), Stage~2 increases recovery under permissive overlap and, more importantly, delivers large absolute gains that remain substantial at strict coverage thresholds (80--100\%), indicating improved boundary completeness and containment rather than fragmented partial overlaps.
The per-species heatmap (Figure~\ref{fig:stage2_cov}c) shows the same trend across genomes: Stage~1 often degrades sharply under strict thresholds, while Stage~2 remains high for most species. These gains specifically indicate improved boundary completeness rather than merely increased neighborhood overlap, which is the operational requirement for actionable experimental follow-up.

To complement coverage-based recovery, we further examine per-species candidate confidence statistics.
Stage~2 increases the mean confidence of predicted loci from 0.733 to 0.788 on average across species, and modestly raises peak locus scores from 0.904 to 0.932; several species exhibit especially large mean-confidence gains (e.g., \textit{Zea mays}: 0.659$\rightarrow$0.795), consistent with strict-coverage recovery improvements.
Distributional analysis also shows a clear right-shift after Stage~2, indicating preferential retention of higher-confidence loci.
In addition, per-species recovery deltas under both overlap-based and strict coverage criteria confirm that improvements are broadly distributed across species rather than dominated by a small subset of genomes.

\begin{table}[t]
\centering
\small
\caption{\textbf{Incremental contribution of each stage.} Each stage is evaluated using the metric aligned with its objective.}
\label{tab:ablation_summary}
\begin{tabular}{p{1.2cm} p{3.0cm} p{2.8cm}}
\toprule
\textbf{Stage} & \textbf{Primary objective} & \textbf{Key result} \\
\midrule
Stage~1 &
Learn a transferable microbial BGC scoring function &
ROC-AUC improves from 0.945 to 0.979 under leave-class-out; F1 improves from 0.775 to 0.878 over DeepBGC. \\
\midrule
Stage~2 &
Mitigate microbial to plant domain shift without plant labels &
Known-BGC recovery improves from 76.5\% to 100\% (Any), and from 29.4\% to 67.6\% at 100\% coverage. \\
\midrule
Stage~3 &
Reduce ProxyFP candidates while preserving ranking utility &
Median proxy-primary ratio decreases from 0.35 to 0.16 across species (Wilcoxon $p=1.53\times10^{-5}$). \\
\bottomrule
\end{tabular}
\end{table}

\begin{table*}[!b]
\centering
\small
\setlength{\tabcolsep}{5pt}
\caption{\textbf{Species-level candidate reduction after Stage~3.}
Candidate counts before (\textbf{Orig.}) and after (\textbf{Kept}) Stage~3 refinement are shown for each species; \textbf{Removed} and \textbf{\% Red.} summarize the reduction in experimental workload.}
\label{tab:stage3_scaling_summary}
\begin{tabular}{lrrrrccc}
\toprule
\textbf{Species} & \textbf{Orig.} & \textbf{Kept} & \textbf{Removed} & \textbf{\% Red.} &
\textbf{Scale$_{\mathrm{prim}}$} & \textbf{Scale$_{\mathrm{mixed}}$} & \textbf{Scale$_{\mathrm{review}}$} \\
\midrule
\textit{Arabidopsis thaliana} & 125 & 79  & 46  & 36.8 & 0.21 & 0.21 & 0.75 \\
\textit{Brassica oleracea}    & 295 & 184 & 111 & 37.6 & 0.21 & 0.22 & 0.75 \\
\textit{Capsella rubella}     & 168 & 112 & 56  & 33.3 & 0.36 & 0.20 & 0.75 \\
\textit{Cucumis sativus}      & 136 & 89  & 47  & 34.6 & 0.35 & 0.19 & 0.77 \\
\textit{Hordeum vulgare}      & 270 & 161 & 109 & 40.4 & 0.27 & 0.20 & 0.71 \\
\textit{Lotus japonicus}      & 191 & 128 & 63  & 33.0 & 0.36 & 0.20 & 0.76 \\
\textit{Nicotiana tabacum}    & 450 & 270 & 180 & 40.0 & 0.25 & 0.18 & 0.71 \\
\textit{Oryza sativa}         & 227 & 141 & 86  & 37.9 & 0.28 & 0.24 & 0.74 \\
\textit{Papaver somniferum}   & 399 & 256 & 143 & 35.8 & 0.37 & 0.24 & 0.72 \\
\textit{Ricinus communis}     & 139 & 91  & 48  & 34.5 & 0.27 & 0.21 & 0.80 \\
\textit{Salvia splendens}     & 361 & 222 & 139 & 38.5 & 0.20 & 0.23 & 0.73 \\
\textit{Sorghum bicolor}      & 373 & 244 & 129 & 34.6 & 0.30 & 0.21 & 0.76 \\
\textit{Solanum lycopersicum} & 203 & 128 & 75  & 36.9 & 0.35 & 0.23 & 0.74 \\
\textit{Solanum tuberosum}    & 243 & 157 & 86  & 35.4 & 0.27 & 0.22 & 0.78 \\
\textit{Taxus chinensis}      & 184 & 114 & 70  & 38.0 & 0.24 & 0.22 & 0.75 \\
\textit{Zea mays}             & 209 & 148 & 61  & 29.2 & 0.38 & 0.24 & 0.79 \\
\bottomrule
\end{tabular}
\vspace{-2mm}
\end{table*}

Together, these results support \textbf{Q2}: MLM-based label-free adaptation systematically mitigates microbial-to-plant domain shift and yields consistent boundary-level gains without relying on plant BGC labels. Since no prior plant-specific ML detector is available, these results also establish the first evidence that a learned BGC scoring model can be successfully transferred and adapted to the plant domain, beyond the capabilities of existing systems.

\vspace{-1mm}
\subsection{Stage 3: GO/KEGG weak supervised learning}
\label{sec:stage3}
Stage~3 addresses \textbf{Q3}: after label-free adaptation, whether GO/KEGG-derived weak supervision can reduce ProxyFP while preserving the ranking utility needed for prioritization.
This step is necessary because in plants, gene-dense primary-metabolism regions can still receive high BGC-likeness scores under domain shift, and comprehensive plant ground truth is unavailable for direct FP measurement.
We therefore use GO\citep{ashburner2000go} and KEGG\citep{kanehisa2000kegg} as functional weak supervision to construct conservative proxy labels and train confidence-weighted soft negatives, targeting primary-like loci without collapsing sensitivity to secondary/mixed signals.
Table~\ref{tab:ablation_summary} summarizes the incremental contribution of each stage. Stage~3 does not further aim to increase known-BGC recovery, but instead refines the adapted score distribution to suppress primary-metabolism-like loci that remain after Stage~2.

Let $r$ denote the primary-like ratio among predicted candidates.
We report $r$ both (i) over all candidates (global calibration) and (ii) over top-ranked candidates.
We quantify the improvement by:
\begin{equation}
\text{ProxyFP\_reduction}(\%) = 100 \times \frac{r_{\text{Stage2}} - r_{\text{Stage3}}}{r_{\text{Stage2}}}.
\label{eq:proxy_fp_reduction_stage3}
\end{equation}

\begin{table}[t]
\centering
\small
\setlength{\tabcolsep}{4pt}
\caption{\textbf{Stage~3 reduces ProxyFP across GO and KEGG evidence.}
We report the number of supported candidate loci and the fraction classified as primary-like before and after Stage~3. ProxyFP reduction is the relative decrease in primary-like ratio from Stage~2 to Stage~3.}
\label{tab:stage2_stage3_summary}
\begin{tabular}{llrrr}
\toprule
\textbf{Evidence} & \textbf{Stage} & \textbf{Total} & \textbf{Prim.-like} & \textbf{Ratio (\%)} \\
\midrule
\multirow{3}{*}{GO} 
  & Stage2 & 4243 & 926 & 21.82 \\
  & Stage3 & 2868 & 323 & 11.26 \\
  & ProxyFP reduc. & \multicolumn{3}{r}{48.40} \\
\midrule
\multirow{3}{*}{KEGG} 
  & Stage2 & 3730 & 1370 & 36.73 \\
  & Stage3 & 2037 & 410  & 20.13 \\
  & ProxyFP reduc. & \multicolumn{3}{r}{45.20} \\
\bottomrule
\end{tabular}
\vspace{-1mm}
\end{table}

Stage~3 fine-tunes the Stage~2 checkpoint using GO/KEGG-derived primary-like evidence as soft negatives.
We assign each predicted locus a primary-like confidence from GO proxy labels and KEGG KO-based primary-metabolism signatures; loci supported by both resources receive an agreement boost (Eq.~\ref{eq:q_fusion}).
The locus confidence is mapped to token-level soft targets (Eq.~\ref{eq:soft_target_final}), and we optimize the confidence-weighted soft-label loss (Eq.~\ref{eq:soft_bce}) jointly with the microbial supervised objective (Eq.~\ref{eq:bce_sup}) under the same architecture and optimization protocol as Stage~2.

\begin{figure}[t]
\centering
\includegraphics[width=\columnwidth]{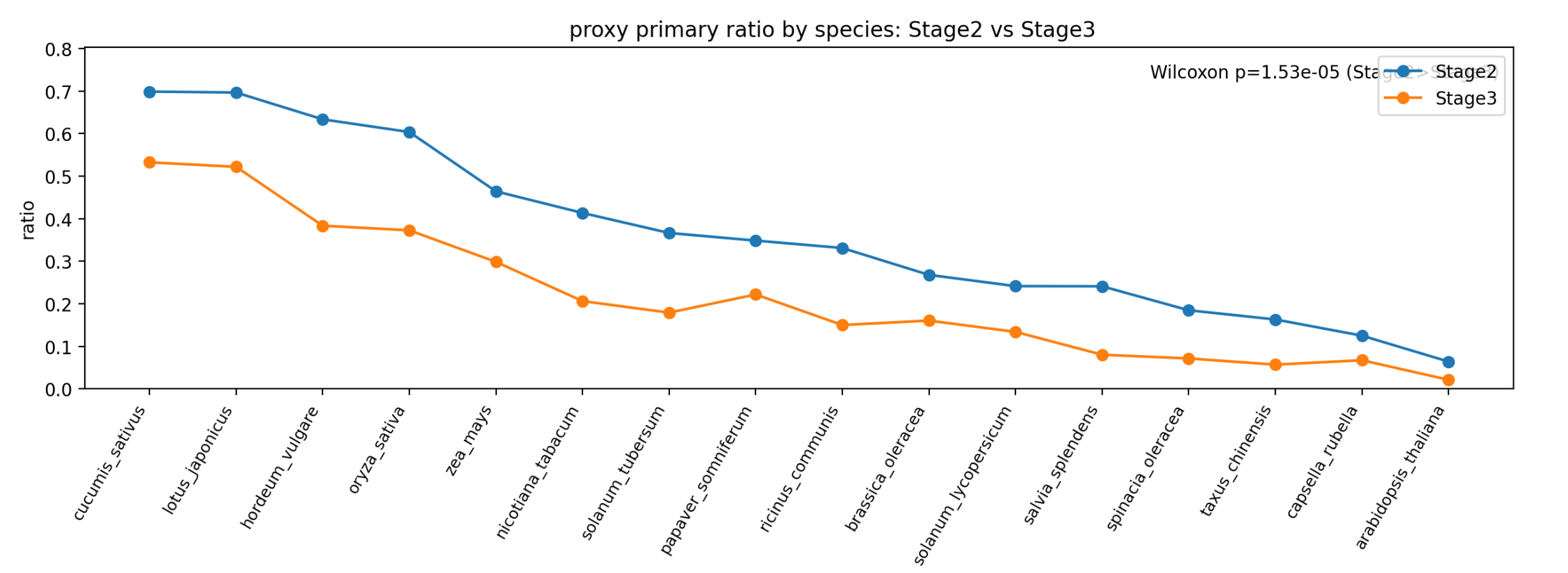}
\vspace{-3mm}
\caption{\textbf{Stage~3 reduces proxy primary-metabolism burden across plant species.}
Each point shows the per-species proxy primary ratio, defined as the fraction of candidate loci labeled as primary-like by GO/KEGG evidence, for Stage~2 and Stage~3. Species are ordered by the Stage~2 ratio. Stage~3 consistently lowers the proxy primary ratio; the annotated $p$-value is from a paired Wilcoxon signed-rank test.}
\label{fig:proxy_primary_ratio_compare}
\vspace{-5mm}
\end{figure}
Table~\ref{tab:stage2_stage3_summary} quantifies the impact of weak supervision on proxy primary-like burden.
Stage~3 reduces the primary-like ratio from 21.82\% to 11.26\% under GO evidence (48.40\% relative reduction) and from 36.73\% to 20.13\% under KEGG evidence (45.20\% relative reduction), indicating that GO/KEGG-weighted soft negatives suppress primary-metabolism-like predictions without requiring plant BGC ground truth.
We report fewer KEGG-supported loci than GO-supported loci mainly due to KO identifier mismatches for RefSeq CDS during KO assignment.

Figure~\ref{fig:proxy_primary_ratio_compare} shows the per-species proxy primary ratio for Stage~2 and Stage~3.
Stage~3 consistently lowers the proxy primary ratio across species, and the paired Wilcoxon signed-rank test confirms the reduction is statistically significant ($p=1.53\times10^{-5}$), suggesting the gain is not driven by a small subset of genomes.

\begin{figure*}[!b]
  \centering
  \includegraphics[width=\textwidth]{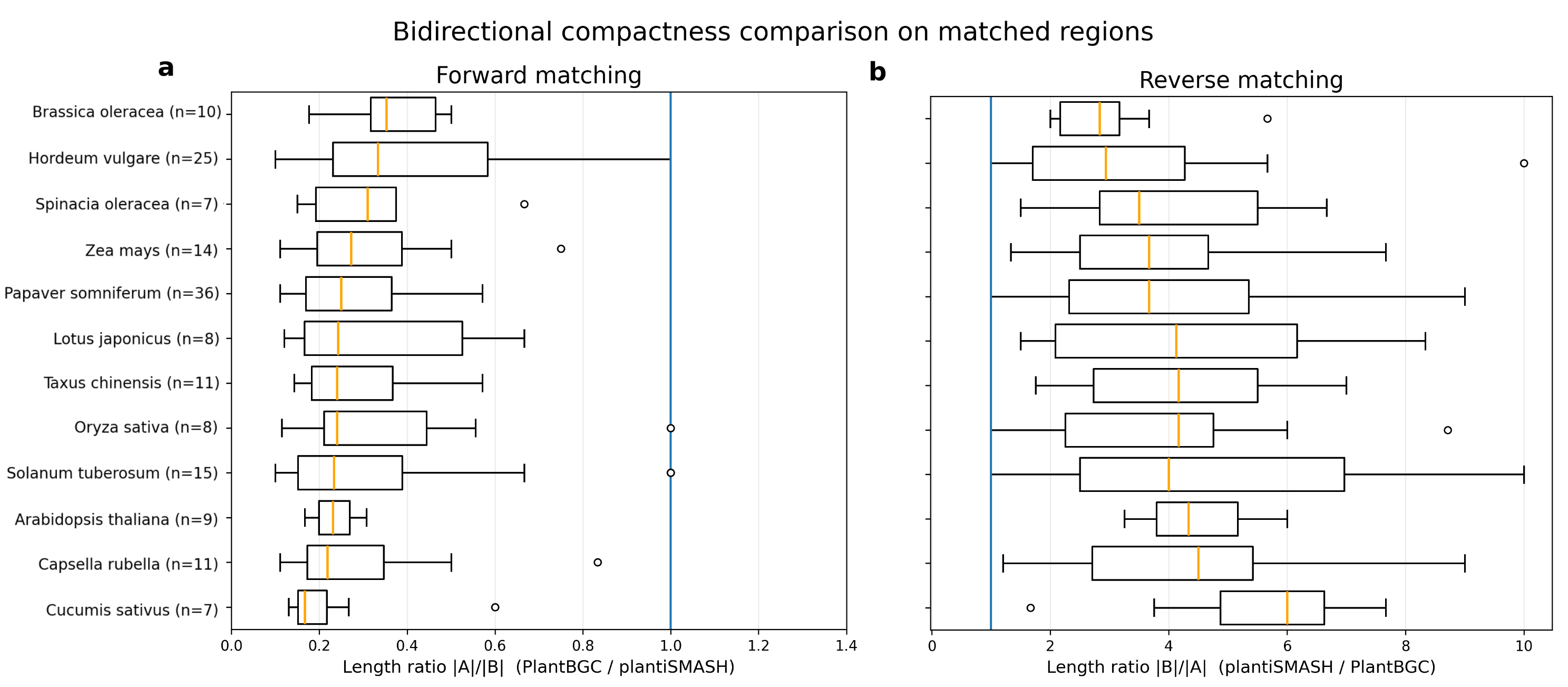}
  \vspace{-4mm}
\caption{\textbf{Bidirectional compactness comparison between PlantBGC and plantiSMASH.}
Pairs are defined by best matches with Prot-IoU $\ge 0.1$.
(a) Forward matching: length ratio $|A|/|B|$ for PlantBGC candidates relative to their matched plantiSMASH clusters; values below 1 indicate more compact PlantBGC loci.
(b) Reverse matching: length ratio $|B|/|A|$ for plantiSMASH clusters relative to their matched PlantBGC loci; values above 1 indicate longer plantiSMASH loci.
Consistent trends in both directions indicate that the compactness difference is not caused by one-directional matching bias such as cluster splitting or merging.
Boxes show the interquartile range and median; whiskers extend to 1.5$\times$IQR, and points denote outliers. Sample sizes ($n$) are shown per species.}
  \label{fig:planti_compact}
\end{figure*}

Table~\ref{tab:stage3_scaling_summary} further summarizes the species-level candidate counts before and after Stage~3 refinement. Across genomes, Stage~3 removes approximately 30--45\% of candidate loci while retaining a focused set of high-priority predictions (e.g., \textit{Nicotiana tabacum}: 450$\rightarrow$270; \textit{Papaver somniferum}: 399$\rightarrow$256; \textit{Salvia splendens}: 361$\rightarrow$222). This indicates that the reduction in proxy primary-like burden translates into a substantially smaller downstream validation workload, while still preserving a nontrivial set of candidate loci for experimental follow-up. Because no prior ML-based plant BGC detector exists, these species-level counts also provide the first practical estimate of the experimental workload associated with a learned plant-domain BGC prioritization framework.

Stage~3 removes primary-like contamination while preserving secondary/mixed enrichment among top-ranked candidates, aligning model outputs with practical screening needs.
Detailed GO/KEGG-based analyses show consistent shifts in label distributions and overlap patterns after Stage~3, including reduced primary-like assignments and stable or increased concentration of secondary/mixed evidence among top-ranked loci.
These trends indicate that Stage~3 acts as a targeted refinement step: it preferentially down-weights primary-metabolism-like predictions while maintaining the enrichment of secondary/mixed signals, which is critical for practical prioritization.
At the species level, the calibrated model consistently reduces the number of primary-like candidates while retaining loci with potential biosynthetic relevance, providing a transparent view of the remaining experimental workload across genomes.

\vspace{-1mm}
\subsection{PlantBGC vs.\ plantiSMASH}
\label{subsec:stage4_planti_compare}
PlantiSMASH\citep{kautsar2017plantismash} is a widely used rule-based plant BGC tool that nominates clusters via curated signatures and heuristic locus rules. 
In contrast, PlantBGC produces a learned, transferable BGC-likeness score track and then aggregates it into gap-free loci, explicitly optimizing for compact, actionable boundaries rather than signature completeness alone.
We therefore perform a head-to-head comparison focused on boundary extent on matched genomic neighborhoods, since boundary inflation directly translates to experimental workload.

We convert outputs of both methods into base-pair intervals on the same reference assembly,
$\bigl(\mathrm{chr}, \mathrm{start}, \mathrm{end}\bigr)$.
Let $A_i$ be a PlantBGC candidate and $B_j$ a plantiSMASH cluster.
For each $A_i$, we match the best-overlap plantiSMASH cluster
$j^{\ast}(i)=\arg\max_j \mathrm{IoU}_{\mathrm{prot}}(A_i,B_j)$, where
$\mathrm{IoU}_{\mathrm{prot}}(A,B)=\frac{|A\cap B|}{|A\cup B|}$.

To compare boundary extent on matched regions, we analyze only pairs with
$\mathrm{IoU}_{\mathrm{prot}}\ge 0.1$ and compute the length ratio
$r=\frac{|A_i|}{|B_{j^{\ast}(i)}|}$.
Ratios $r<1$ indicate PlantBGC predicts a shorter (more compact) locus than plantiSMASH for the same matched genomic region.

To control for possible bias when one method splits or merges loci, we also perform the reverse comparison.
For each plantiSMASH cluster $B_j$, we identify its best-overlap PlantBGC locus
$i^{\ast}(j)=\arg\max_i \mathrm{IoU}_{\mathrm{prot}}(A_i,B_j)$ and compute
$
r_{\mathrm{rev}}=\frac{|B_j|}{|A_{i^{\ast}(j)}|}.
$
In this direction, values $r_{\mathrm{rev}}>1$ indicate that plantiSMASH predicts a larger locus than PlantBGC on the same matched region.

Fig.~\ref{fig:planti_compact} shows that PlantBGC loci are consistently more compact than plantiSMASH under both matching directions.
Across all forward matches, the pooled median ratio is 0.278 and 93.8\% of matched pairs satisfy $|A|<|B|$.
Conversely, across reverse matches, the pooled median ratio is 3.92 and 96.1\% of matched pairs satisfy $|B|>|A|$.
The same trend is observed across nearly all species, indicating that the compactness result is not an artifact of one-directional matching (e.g., cluster splitting or merging).
This suggests that when both methods identify the same genomic neighborhood, PlantBGC more often delineates a narrower locus while plantiSMASH tends to include substantially larger flanking regions.
Such compact predictions are advantageous for downstream prioritization and experimental validation because they reduce the number of neighboring genes that must be inspected or functionally tested.
In addition to compactness, this comparison is intended to reflect boundary precision and the resulting experimental workload, rather than to claim overall superiority across all aspects of BGC detection. Despite these boundary differences, overall agreement between methods remains substantial.
Best-match protein-overlap IoU is moderate overall (pooled median IoU $=0.148$, maximum $0.805$; 61.7\%/31.6\%/9.9\% of pairs exceed IoU thresholds 0.1/0.25/0.5), indicating that the two methods frequently nominate the same genomic neighborhoods but disagree on the precise locus boundaries.

\section{Contributions}
We present PlantBGC, to our knowledge the first Transformer-based framework that transfers BGC supervision from microbes to plants while minimizing reliance on plant BGC labels. PlantBGC learns a transferable token-level scoring function from curated microbial BGCs, performs label-free microbial-to-plant adaptation via masked language modeling on unlabeled plant Pfam sequences to mitigate domain shift and improve strict coverage-based recovery of curated plant BGCs, and further applies GO/KEGG-derived weak supervision as confidence-weighted soft negatives to calibrate predictions and reduce ProxyFP across species. On matched genomic regions, PlantBGC produces substantially more compact loci than plantiSMASH, directly lowering downstream validation effort. Future work will incorporate additional coordinate-resolved plant BGCs, integrate orthogonal evidence sources, and experimentally validate top-ranked Stage~3 candidates.

% ------------------------------------------------------------------------
% End section 2
% -----------------------------------------------------------------------

\bibliographystyle{ACM-Reference-Format}
\bibliography{reference}

%%
%% If your work has an appendix, this is the place to put it.

\end{document}